# Breaking the Vault: A Case Study of the 2022 LastPass Data Breach

Jessica Gentles, Mason Fields, Garrett Goodman, Suman Bhunia
*Department of Computer Science and Software Engineering, Miami University, Oxford, Ohio, USA 45056*
Email: gentlejc@miamioh.edu, fieldsmn@miamioh.edu, goodmag@miamioh.edu, bhunias@miamioh.edu

*Abstract*—Managing the security of employee work computers has become increasingly important as today's work model shifts to remote and hybrid work plans. In this paper, we explore the recent 2022 LastPass data breach, in which the attacker obtained sensitive customer data by exploiting a software vulnerability on a DevSecOps engineer's computer. We discuss the methodology of the attacker as well as the impact this incident had on LastPass and its customers. Next, we expand upon the impact the breach had on LastPass as well as its customers. From this, we propose solutions for preparing for and mitigating similar attacks in the future. The aim of this paper is to shed light on the LastPass incident and provide methods for companies to secure their employee base, both nationally and internationally. With a strong security structure, companies can vastly reduce the chances of falling victim to a similar attack.

*Index Terms*—remote code execution, keylogger, database security, cloud security, data breach, LastPass

## 1. Introduction

In August 2022, LastPass found suspicious activity from one of their software engineers in a cloud-based development environment. After reviewing the logs, they found that an unidentified person or group had stolen the engineer's credentials to access this environment and launched a full investigation [1]. However, they were unaware that this was only the beginning of the attack. Before LastPass was able to reset credentials and lock the threat actor out, they were able to extract highly sensitive data. The stolen customer data included encrypted password vaults, customer usernames, customer addresses, and customer email addresses. Moreover, they stole fourteen development source code repositories and other data from LastPass, including API keys and encrypted credentials.

Password managers like LastPass are large targets for cyberattacks due to the vast amounts of incredibly sensitive customer data they contain. As Fig. 1 shows, LastPass sells itself as a vault for more than just usernames and passwords; they openly encourage customers to save credit cards, social security numbers, driver's license numbers, and work credentials in their vaults [2]. Seeing that Last-Pass stores more than simply usernames and passwords, it is clear that a hacker would be highly motivated to breach their database.

The identity of the attacker in this breach is still unknown, but it is clear from their methods that they are experienced as they meticulously targeted their entry point into LastPass' database backups. This data breach was sparked by the threat actor targeting a software engineer's computer to steal their credentials, leading to access to the development environment and cloud backups of the database. The vulnerabilities in a third-party media app on the engineer's computer allowed for remote code execution, which enabled the attacker to install a keylogger on their device. LastPass failed to implement strong security policies for employees working from their home network, eventually leading to this data breach. When handling highly sensitive customer and business information like this, security has to be the highest priority. It is also concerning for an employee of the company to be running vulnerable third-party software on a machine holding credentials that give access to customer data.

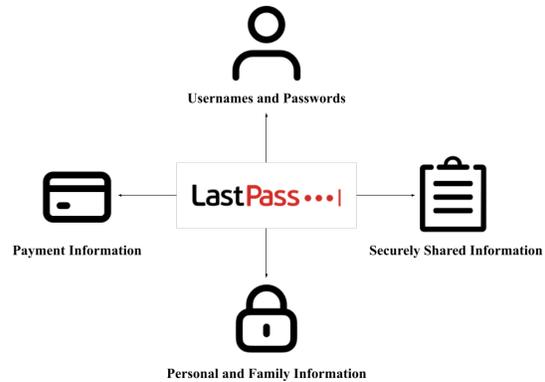

Figure 1. A diagram of the data that LastPass customers store in their password vaults.

Most of the critical data that was stolen (data from the password vault) is encrypted, and LastPass claims that it can only be decrypted with a user's master password, which was not stolen by the attacker. Regardless, an attack of this caliber is unacceptable if you are a customer or business using LastPass' services. This breach has severely damaged LastPass' reputation, and many customers have lost trust as a result. Since the breach, LastPass has made significant investments in security, including a restructuring of security policies. They also plan to expand the use of encryption in their backup infrastructure and applications as a whole. With these policies implemented, LastPass may regain the trust of their customers and provide truly secure storage for user passwords and private information.

The remainder of this paper is organized as follows. Section 2 introduces LastPass as well as software that was relevant to the breach. It also introduces what is known about the attacker. Section 3 dives into the methodology behind the attack, including the vulnerabilities exploited and how the attacker found sensitive information after gaining access. Section 4 details the impact this breach had on LastPass and its customers. In Section 5, we suggest defense solutions to prepare for an attack of this nature in the future. Lastly, Section 6 provides a summary of the key points of this breach and the suggestions we made to enhance security.

## 2. Background

Before presenting the attack itself, it is crucial to understand the background of the victim, LastPass, and the attacker.

### 2.1. LastPass: Background and Software Utilized

This section discusses the background of LastPass, its business model, and the software the company uses. It also provides some insight into the attackers. However, their identity is still unknown.

Password managers like LastPass help individual customers and companies to centralize their passwords in a secure location. This requires customers to only remember a single password (that is, their LastPass master password), and LastPass will take care of their logins for all of their accounts. Furthermore, they allow for the storage of sensitive information such as social security and credit card numbers. LastPass is the most popular password manager on the market, providing services to over 30 million customers and over 100,000 businesses [3]. Most other account-based online services will store basic Personal Identifiable Information (PII) to keep track of their users, but password managers like LastPass are a different story; they serve as a secure storage place for user logins to all of their services. Considering the sensitivity of these data and the sheer volume of data LastPass is housing, it is critical to keep it secure.

LastPass uses Amazon Web Services (AWS) for cloud storage and development. In particular, they use Amazon Simple Storage Service (Amazon S3), which is an object storage service designed to store incredibly large amounts of data while offering scalability, data availability, security, and performance. They also provide an advanced security feature that alerts owners of suspicious activity within the database [4]. Supply chain vulnerabilities remain one of the largest cybersecurity threats, as demonstrated in the SolarWinds Orion Platform breach [5]. The LastPass breach highlights how attackers exploited weaknesses in third-party software to implant keyloggers and gain unauthorized access to sensitive credentials.

### 2.2. Mandiant: Incident Response Firm

Mandiant is considered to be the cyber security market leader in threat intelligence, being employed by enterprises, governments, and law enforcement [6]. They offer security services for threat intelligence and detection, as well as consultation to improve company security or clean up an incident. In this case, they were called in to help contain, eradicate, and recover following the breach. The extent of Mandiant's contribution to cleaning up the breach has not been disclosed.

### 2.3. The Perpetrator of the Breach (Unknown)

The exact identity of the threat actor that attacked LastPass is still unknown. What is known is that it was an intricate and sophisticated attack that required a significant amount of planning and coordination. For example, they gained access to LastPass' systems by targeting the home computer of one of the four senior DevSecOps engineers. The threat actor then exploited third-party software to implant a keylogger and gain access to their cloud database [7]. Considering the planning and reconnaissance that had to go into this attack, this means that it was likely an experienced group. We recommend you search for the identity of the attacker yourself, as it may have been discovered since the writing of this paper.

## 3. Attack Methodology

This section discusses the attack methodology behind the LastPass breach, including the steps from the attacker's initial reconnaissance, what they did following the data extraction, and everything in between. Fig. 2 shows the entire timeline of the breach. Here, we will discuss the weak points the attacker used to gain access, the tools the attacker used to infiltrate the system, and the vulnerabilities that were exploited.

### 3.1. Scanning and Reconnaissance of LastPass's Infrastructure

It is impossible to know the exact software utilized by the threat actor for their reconnaissance, and we do not know what LastPass assets they tried to exploit as well, but we do know the entry point they eventually decided upon, a DevSecOps' home computer. Through the threat actor's likely rigorous scanning and enumeration phase, they came to the conclusion that attacking a DevSecOps home computer would be their best bet at gaining access to LastPass data. The threat actor found that there were four DevSecOps engineers who had access to the corporate vault containing decryption keys [10]. Rather than attempting to steal lower-level credentials and escalating privileges, gaining access to a DevSecOps engineer's credentials would allow for immediate access to the data they were looking for. Additionally, if they could obtain credentials without being caught immediately, they would have plenty of time to locate and steal the data they were after.

With this in mind, the threat actor found that third-party media software running on a DevSecOps engineer's home computer had a fatal vulnerability; once exploited, they would be able to remotely execute code on it. The threat actor would later take advantage of this detrimental vulnerability to steal the engineer's credentials.

### 3.2. Accessing LastPass Data

With this exploit, the threat actor was able to install a keylogger on the engineer's computer. Using the keylogger, they were able to capture the engineer's master password as it was entered after logging in with multifactor authentication [11]. The bulleted list below summarizes the steps on how the threat actor exploited a vulnerability in the engineer's computer to steal credentials.

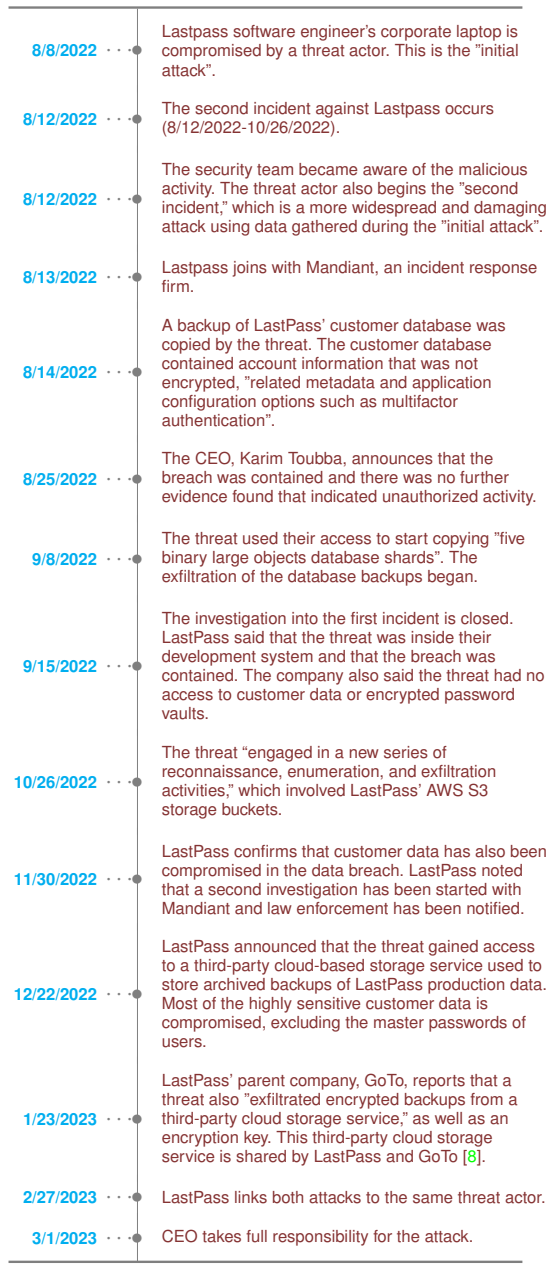

Figure 2. Timeline of the LastPass Data Breach [9]

- The attacker exploits the vulnerability via the engineer's computer and the vulnerable software.
- The attacker installs a keylogger.
- Credentials are captured via the keylogger and are compromised.

With the DevSecOps engineer's credentials secured, access to the encrypted corporate vault was easy. Furthermore, the threat actor was able to keep their identity unknown and bypass intrusion detection systems thanks to the use of a third-party VPN service. This hid the origin of their activity when accessing the cloud-based development environment and allowed them to impersonate the software engineer [1].

With the system infiltrated, the threat actor exfiltrated fourteen code repositories from the cloud-based development environment. As disclosed by LastPass, "*some of these source code repositories included cleartext embedded credentials, stored digital certificates related to our development environments, and some encrypted credentials used for production capabilities such as backup* [1]." Thankfully, the decryption keys for these credentials were not available to the software engineer that the threat actor was impersonating. More importantly, no customer data was stolen at this point, as neither customer nor vault data are stored in the development environment. At the time, it seemed that the incident had been contained and the damage had been mitigated. However, the true attack and data exfiltration were just about to begin.

The LastPass breach highlights the risks associated with exposed API keys and embedded credentials in source code. Similar vulnerabilities have been exploited in previous API-based data breaches, such as the LinkedIn API scraping incident [12] and the Parler data breach [13], where attackers leveraged weak authentication controls to exfiltrate sensitive user data.

### 3.3. Leveraging The Stolen Data

Before the credentials and other information obtained by the threat actor could be reset, the threat actor accessed a shared cloud storage environment powered by AWS. LastPass was completely oblivious to this until they were alerted by AWS GuardDuty. AWS GuardDuty is a service offered by Amazon to help monitor AWS accounts and data with machine learning and integrated threat intelligence. It continuously monitors all AWS accounts for potentially malicious activity and documents its findings [14]. By the time LastPass was alerted of this, the threat actor had already extracted customer and vault data. LastPass stands by the fact that vault data are encrypted with a unique key generated based on each user's master password, but knowing that someone accessed this information is still incredibly concerning for all customers and businesses [15]. Moreover, cases of cryptocurrency theft have been linked to the LastPass breach. This will be discussed in more detail in a later section, but multiple LastPass users who stored their wallet keys in their vault reported having crypto siphoned out of their wallets following the breach. However, this incident is currently being investigated as of the writing of this paper. This brings into question whether or not all of the vault data was encrypted or if the attacker was able to decrypt it. Despite the confidentiality of the stolen vault data, there was also a significant amount of PII about LastPass customers and businesses that was stolen. These stolen data are summarized in Table 1. This data can easily be sold for profit to data brokers or elsewhere on the dark web, putting customers at risk.

Once the threat actor gained access to a DevSecOps engineer's credentials, they were able to infiltrate LastPass' shared cloud storage environment. This type of credential theft and subsequent session hijacking has been previously analyzed in cases such as the CVS Health Data Breach [16], where attackers leveraged stolen login credentials to impersonate legitimate users.

TABLE 1. A SUMMARY OF THE INFORMATION EXTRACTED BY THE THREAT ACTOR (BOTH FROM LASTPASS AND THEIR CUSTOMERS).

| Stolen Information | Details/Potential Impact |
|---|---|
| Password Vaults | - Encrypted<br>- Potential of accessing all customer accounts based on strength of their master password |
| Customer Information | - Unencrypted<br>- Usernames<br>- Billing addresses<br>- Email addresses |
| Vault Metadata | - Name of folders and websites that customers use LastPass to store logins for |
| Development Data | - Source code<br>- API keys |

## 4. Impact

While the attack occurred over a year ago, the LastPass incident continues to have long-lasting effects on both small-scale and large-scale customers.

### 4.1. Warnings to Switch Password Managers

After the LastPass incident, many influential blogs, such as Forbes and CNET, made clear recommendations to switch away from the LastPass password manager [17], [18]. Some suggestions they made for users were to switch to Bitwarden, 1Password, Keeper, NordPass, and Dashlane. Before the data breach, in 2020 and 2021, LastPass was regarded as one of the go-to password managers and notably made the list of the top 5 password managers on FRSECURE, Security Trails, IGN News, and Cyber Tec Security [19]–[22]. LastPass was also recognized as the "best password manager overall" by Business Insider in 2020 [23]. Davey Winder, a veteran cybersecurity and tech analyst, Forbes journalist, and hacker, noted that the LastPass data breach is not the main reason why he would recommend users to switch password managers, as "absolute security is a complete fallacy" [17]. He stated that LastPass' lack of clear communication with customers about the incident is one of his main reasons for recommending the change. Following the data breach, many websites that previously supported LastPass, such as Wired and CNET, have now removed LastPass from their lists of the top password managers due to the data breaches [18], [24]. While CNET stated that they would "conduct a thorough re-review of the service in the future, after which [they] will reevaluate whether LastPass should return to [their] list of best password managers," it is clear that the data breaches have left lasting reputational damage to the LastPass brand.

### 4.2. Financial Loss For Customers

The financial impact of the LastPass breach, particularly the loss of cryptocurrency, mirrors trends seen in other high-profile ransomware attacks targeting critical infrastructure. For example, the ransomware attack on the DC Metropolitan Police Department [25] demonstrated how stolen sensitive data can be used for extortion and financial gain. Most timelines surrounding the LastPass data breach list March 1st of 2023 as the last major event in this data breach. March 1st of 2023 is when Karim Toubba announced that he took full responsibility for the data breach that occurred. Although it is widely known that customer data was compromised during the LastPass data breach, it is not as widely known that the data breach led to a loss of 4.4 million dollars worth of cryptocurrency on October 25th, 2023. This 4.4 million dollars lost is only a small percentage of the 35 million dollars total that had been stolen as a result of the data breach [26]. According to The Verge, over 150 LastPass customers experienced cryptocurrency theft in the months following the data breach [27]. The technology website also reported that "two to five high-value [cryptocurrency] heists" occurred every month between December 2022 and September 2023, following the data breach [27]. Taylor Monahan, lead project at Metamask, a popular cryptocurrency wallet software, stated that "the common thread among nearly every victim [of these cryptocurrency attacks] was that they'd previously used LastPass to store their seed phrase," which is the private key used to unlock cryptocurrency investments [28].

### 4.3. Awareness About and Conflicting Opinions About Password Managers

While the LastPass data breach has been contained, many Americans remain apprehensive about password managers. According to a survey done by the Password Manager, "65% of Americans do not trust password managers" [29]. Password Manager surveys also revealed that nearly half of Americans believe that "nothing could motivate them to use [a password manager] in the future" [29].

Multi-vector cyberattacks have become increasingly sophisticated, leveraging various attack surfaces such as credential theft, session hijacking, and API abuse. A similar approach was observed in the Astoria data breach [30] and the Accellion File Transfer Appliance ransomware attack [31], where attackers combined multiple tactics to maximize data exfiltration.

On the other hand, according to PCMag and as shown in Fig. 3, more Americans are using password managers than ever before [32]. This shows conflicting emotions regarding password managers. While data breaches are alarming when they occur, on average, password managers make users less susceptible to password-based attacks because they eliminate the need for users to recall passwords and force them to develop stronger passwords due to length and character requirements [24]. In the last decade, the most popular passwords remained "123456" and "password," even though 85% of Americans are aware of the dangers of using the same password across websites [24], [29]. This shows a growing need for many people to use password managers or a better password management and creation system.

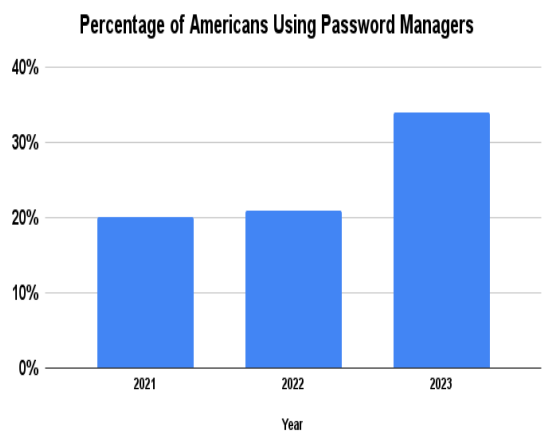

Figure 3. A chart showing the increasing usage of password managers in the United States [29].

## 5. Defense Solution

While absolute security is not possible, there are many actions that customers, as well as businesses, can take to ensure their systems are more secure.

### 5.1. LastPass Defense Solutions For Free, Premium, and Family Customers

For free, premium, and family customers, LastPass recommends making changes to their master passwords, turning on dark web monitoring, and ensuring that there is "multifactor authentication (MFA) for [their] vault" [33]. The first thing LastPass lists in their security bulletin is to create a strong and unique master password for customer accounts. This includes using a minimum of 12 characters, with a preference for passwords even longer than 12 characters, with at least one of each: upper case, lower case, numeric, symbol, and special character. LastPass also emphasizes that customers need to make sure that their master password is not used anywhere else on the internet as that makes the password more vulnerable to breaches. As the LastPass master password gives access to all of the customer's other passwords, it is absolutely essential that the master password is fully unique. LastPass also suggests that customers do not use any personal information as this may make it easier for threats to guess user passwords. Finally, LastPass suggests using a randomly generated master password for your master key as these are far less vulnerable to being cracked by threats. LastPass suggests that customers use their LastPass Password Generator but any randomly generated password will be stronger than a password created by an individual [33]. While LastPass aims these suggestions at their customer base, these suggestions are also applicable to any password on any system, as they are the current suggestions to increase the strength of the password.

LastPass also suggests that customers review their passwords for overall password strength using their security dashboard. Within this dashboard, customers can view scores for all passwords. The password manager application company suggests that customers immediately change any passwords that are classified as unsafe by their system. LastPass suggests that users turn on dark web monitoring as this feature "evaluates email addresses saved in your vault items" and alerts customers immediately "if any of [their] email addresses have been found in the database of credentials breached in third-party security incidents" [33]. If customers are alerted using the dark web monitoring feature, LastPass will guide them through the process of making sure their passwords are changed for every site associated with third-party breaches [33].

LastPass' last main suggestion is that customers use multifactor authentication. Multifactor authentication ensures that customers take a second step in order to gain access to their accounts. This helps add extra security to their LastPass accounts. They also encourage customers to utilize multifactor authentication for their LastPass vault. If customers have already added multifactor authentication to their accounts, they are encouraged to go a step further and regenerate their multifactor authentication secret [33]. With these suggestions, LastPass customers will have a reduced risk of password breaches. Table 2 summarizes these recommendations.

TABLE 2. A TABLE SHOWING CURRENT LASTPASS RECOMMENDATIONS FOR CUSTOMERS [33].

| Vulnerability | LastPass Recommendations |
| --- | --- |
| Weak master passwords | Reset master passwords (optional). Ensure that you have a strong master password if using a password manager. |
| Reused master password | LastPass strongly suggests that users do not use their password anywhere but LastPass. |
| Review your overall password strength | LastPass recommends reviewing your password strength using their security dashboard. |

### 5.2. LastPass Defense Solutions For Business Administrators

For business administrators, LastPass has added suggestions to ensure optimal security for business accounts. LastPass suggested that, in addition to all the other practices mentioned free, premium, and family customers, businesses should review their "shared folders accessed by users with a low iteration count" report. This feature provides a list of "shared folders that can be accessed by users with a low iteration count" [33]. LastPass also suggests that users review the super admin best practices to make sure that super admins are aware that when they change passwords in the account, they may put the entire account at risk if they fail to ensure the strength of the password.

SIEM Splunk integration customers are also advised to reset their instance token. SCIM, Enterprise API, and SAML customers were also advised to reset their keys. LastPass users are also advised to check their security reports to ensure that they understand the "risk of any exposed URLs and any associated session IDs or parameters stored with these URLs" [33]. Businesses are encouraged to communicate with their customers about those risks and to enable dark web monitoring for their users.

### 5.3. Defense Practices to Prevent Data Breaches

Sections 2 and 3 cover in detail how the LastPass data breach occurred. As stated previously, absolute security is not possible to achieve, but there are some best practices that can be taken to ensure that incidents such as the LastPass data breach are less likely to occur. According to Stack Identity, for the first incident where the threat actor was able to compromise an employee's corporate laptop to access a cloud-based development environment, one way that LastPass could have potentially prevented this incident is by ensuring to continuously monitor the "provisioned access risk and policies associated with an identity" and remove any unessential or excessive access permissions. This would have made it more difficult for a threat to compromise the system [34]. According to PCMag, the software engineer whose corporate laptop was compromised was using an outdated version of the "Plex Media Server software" [35]. This version of the Plex Media Server software had a known vulnerability, CVE-2020-5741, that the threat actor was able to exploit. This vulnerability was patched in 2020, and it was publicly known that there was a vulnerability in this version [35]. It is common practice that one of the best ways to avoid security breaches is to ensure that all software is up-to-date and to stay knowledgeable about any vulnerabilities that may be in any used software.

For the second incident, where the threat actor targeted another employee, a DevOps engineer, LastPass could have potentially kept better "track of permissions [and] policies defined for roles providing cross-account access to federated identities" [34]. The password manager software company also should have generated reports every 24 hours to "monitor access to the Key Management Services and avoid using permanent secret/access keys for any orchestration/backup platforms storing the key-value pair to access the data assets" [34]. While these actions would have made their system less likely to be breached, it is important to note that no system is able to achieve perfect security. However, these suggestions, summarized in Table 3, could have potentially helped LastPass achieve a more secure system.

TABLE 3. A TABLE SHOWING CURRENT RECOMMENDATIONS TO AVOID DATA BREACHES LIKE THE LASTPASS DATA BREACH [34].

| Vulnerability | Recommendations |
|---|---|
| First Incident | - Continuously monitor user permissions.<br>- Keep up-to-date versions of any software. |
| Second Incident | - Keep track of permissions set "for roles providing cross-account access" [34].<br>- Generate reports every 24 hours.<br>- Keep track of who has access. |

### 6. Conclusion

The LastPass data breach serves as a reminder that even well-regarded password management services are not invincible. As the amount of software used grows, it is important to stay cognizant of the problems that may arise when security is not put as a top priority in software systems. Although the stolen vault data remains encrypted, there were many consequences that stemmed from this incident, and the risk of those master passwords being decrypted still remains. The decryption of these passwords could pose a significant risk of an increase in attacks on vulnerable accounts if previous and current users do not follow the suggested steps to take after any data breach. This LastPass data breach highlights the importance of using a layered security approach, where strong master passwords and multifactor authentication are crucial. It also highlights the need for developers and security employees to remain up-to-date on the vulnerabilities in the different systems we use. Users of different password managers should make sure that they understand the risk that comes with the rewards of using a password manager. With this LastPass data breach seemingly contained, it is important to look at the lessons that can be learned from this experience for both users and developers of software systems. Companies must prioritize transparent communication, rigorous security audits, and continuous review of access permissions. Furthermore, all businesses should make a commitment to developing software that puts the security of customers and the business as a top priority in light of the constantly emerging threats. Users of software systems need to ensure that they are educated about the risks of the software systems they interact with and should practice password hygiene when using all software systems. The LastPass breach serves as a reminder for the software industry to prioritize user security and continue to develop solutions that will aid both businesses and customers alike.